\documentclass[10pt, a4paper]{article}
\usepackage{hyperref}
\usepackage{lrec-coling2024} 
\usepackage{multirow}
\title{Enhancing Knowledge Retrieval with Topic Modeling for Knowledge-Grounded Dialogue}

\name{Nhat Tran, Diane Litman} 

\address{University of Pittsburgh\\
         Pittsburgh, PA  15260 USA\\
         nlt26@pitt.edu, dlitman@pitt.edu\\
         }

\abstract{
Knowledge retrieval is one of the major challenges in building a knowledge-grounded dialogue system. A common method is to use a neural retriever with a distributed approximate nearest-neighbor database to quickly find the relevant knowledge sentences. In this work, we propose an approach that utilizes topic modeling on the knowledge base to further improve retrieval accuracy and as a result, improve response generation. 
Additionally, we experiment with a large language model, ChatGPT, to take advantage of the improved retrieval performance to further improve the generation results.
Experimental results on two datasets show that our approach can increase retrieval and generation performance. The results also indicate that ChatGPT is a better response generator for knowledge-grounded dialogue when relevant knowledge is provided.
 \\ \newline \Keywords{knowledge-grounded dialogue, topic modeling, dense retrieval} }

\begin{document}

\maketitleabstract

\section{Introduction}

In knowledge-grounded dialogues, to find relevant knowledge passages from a large knowledge base, retrieval-based approaches use two encoders to encode both dialogue history and the knowledge base into the same vector space.
The encoded dialogue history is treated as an input query to quickly find the relevant knowledge by retrieving the top-K passages in the encoded knowledge base based on a similarity score (e.g., dot product).
Improvement in any of these two encoders can potentially lead to increased performance of knowledge retrieval.

\begin{figure}[t!]
\includegraphics[width=1\columnwidth]{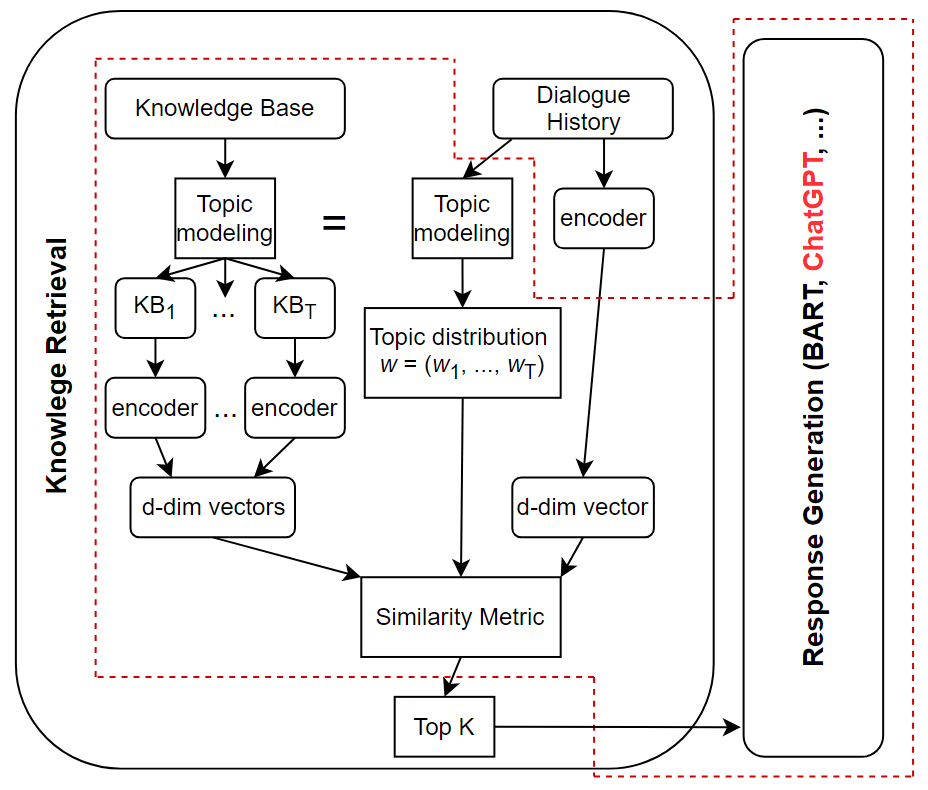}
\caption{The modified retrieve-then-generate framework (based on RAG) with our contribution {\color{red}highlighted}. The two topic modeling modules are  the same one trained on the knowledge base.}
\label{fig:contrib}
\end{figure}

While some prior work focused on improving the dialogue history encoder \cite{our-sigdial}, ours focuses on the knowledge base encoder.
Specifically, we use topic modeling to cluster the knowledge base and train a separate encoder for each cluster. 
We then incorporate the topic distribution of the input query into the similarity score to find the top-K passages. 
Due to impressive performance across various natural language processing (NLP) tasks of large language models (LLMs) such as ChatGPT, we also experiment with using ChatGPT as the response generator, with and without the retrieved knowledge.
Figure \ref{fig:contrib} shows our focus within the retrieve-then-generate framework.

Our contribution is threefold. First, we propose a modification utilizing topic modeling to the widely used RAG (retrieval-augmented generation) model 
and achieve improved performance and verify that using validation sets is a reliable way to pick the optimal number of topics. Second,
we show that combining our approach which manipulates the knowledge base with 
an approach focusing on building a better input query 
can further improve performance. 
Finally, we find that the relevant knowledge is essential for ChatGPT to achieve the best performances. We also make our source code available at
\href{https://github.com/nhattlm95/tm_kg_dialogue}{https://github.com/nhattlm95/tm\_kg\_dialogue}. 

\section{Related Work}
For knowledge-grounded NLP, knowledge retrieval is a crucial step \cite{Eremee-dialogue}.
Although LMs can be embedded with knowledge \citep{lama, lama2, autoprompt, how_much_knowledge}, retrieve-then-generate models still yield higher performances in knowledge-intensive tasks \cite{kilt, wow_dataset, li-knowledge}. 
 Our work follows this line of research, in which a dedicated knowledge retrieval component is used.

 Recent dense retrieval methods \citep{dpr, rag, ance}, which encode text as a latent vector and use their distances to determine the relevance, have outperformed the sparse methods such as TF-IDF or BM25 \cite{bm25}.
In this work, we utilize dense retrieval by modifying the retriever module and the way to calculate the similarity scores of the popular RAG model \cite{rag} with the help of topic modeling.

The concept of \textit{topics} has not been explored much in knowledge-grounded dialogue.
\citet{knowexpert} proposed an end-to-end
framework that uses topic modeling to skip the explicit retrieval process and inject knowledge into the pre-trained language models for knowledge-grounded conversations. \citet{our-sigdial} 
tries to maintain similar `topics' (e.g., turns grounded in the same document) in the dialogue history used as input queries in dense retrieval. Those works are different from ours as we focus on improving the knowledge retrieval component with the help of topic modeling on the knowledge base.

Although ChatGPT \cite{chatgpt} has shown great performance in various NLP tasks 
\cite{chatgpt-bench},
recent works in knowledge-grounded dialogue \cite{li-knowledge, zhao-knowledge, wu-knowledge, gowriraj-language} have not utilized it as a response generator.
Our work tests the potential of ChatGPT to generate responses that need to be grounded in certain knowledge, with the presence/absence of the required knowledge.

\section{Method}
\label{sec:method}

We first perform \textbf{topic modeling} 
to cluster the training knowledge base into a pre-defined number (\textit{\textbf{T}}) of topic clusters. We use the contextual topic model (CTM) from \citet{ctm} which has shown better topic coherence compared to traditional methods. The 
major components of CTM are a neural topic model Neural-ProdLDA \cite{neural_lda} and pre-trained Sentence Transformers embedding \cite{bert_embedding}. Once trained, the 
model can output a T-dimension vector $w =(w_1, w_2, ..., w_T)$ given an input sequence, which is the probability distribution of the pre-clustered topics.

To find the top-K relevant knowledge passages from a large knowledge base for a given dialogue history \textit{H}, we modify  
\textbf{Dense Passage Retrieval (DPR)} \citep{dpr}. Traditionally, it utilizes two BERT encoders \citep{bert}, a document encoder ($BERT_d$) and a query encoder ($BERT_q$), to encode the knowledge passages and the dialogue history to the same d-dimensional space. The document encoding is done offline and indexed in a database such as FAISS \citep{faiss} which can retrieve the top-K at inference time quickly if the relevance score
between two vectors is calculated as their dot product.

However, since we have a T-cluster knowledge base, for each cluster $t_i$, we train a separate document encoder $BERT_d^i$. 
Given the topic distribution of the dialogue history H calculated using CTM as $w =(w_1, w_2, ..., w_T)$, 
to find the top-K passages, we first retrieve the top-K passages from each cluster $t_i$, with the relevant score of a passage $p$ inside the cluster calculated as:
\begin{equation} \label{eq:1}
    BERT_q(H) \cdot BERT_d^i(p) \times w_i 
\end{equation}
where $\cdot$ is dot product and $\times$ is  multiplication.
Then, we choose the top-K from these K $\times$ T retrieved passages. We call this version \textbf{DPR-topic}.

To generate the 
response, we use \textbf{RAG} \citep{rag}. It has a retriever 
(DPR) and a generator module (BART, \citet{bart}).
Given the dialogue history as an input query, the retriever finds the top-K relevant passages, and the generator takes the dialogue history and retrieved top-K  passages 
to generate the response. 
The retriever is non-parametric so any pre-trained model can be used.
We use \textbf{DPR-topic} as the retriever 
and do not touch the RAG query encoder or the generator module. 
Our model is called {\bf RAG-topic}. 

For MultiDoc2Dial data (Section \ref{sec:experiment}), we also experiment with a RAG-based model (RAG-context) that uses an algorithm to select relevant turns in the dialogue history \cite{our-sigdial}. Since it only changes the query of RAG, our approach which manipulates the knowledge base can be applied in the same way we modify the RAG model.
We call this model \textbf{RAG-context-topic}.

Finally, we experiment with {\bf ChatGPT\footnote{We use \href{https://openai.com/research/gpt-4}{GPT-4} from OpenAI.}}. Besides asking ChatGPT to generate the response directly given the dialogue history, we 
feed external knowledge as input to ChatGPT to give it the necessary knowledge. The external knowledge is text retrieved from the retriever (the web page containing the top-1 retrieved passage).



\section{Implementation Details}
For topic modeling, we use CTM\footnote{\url{https://tinyurl.com/3hb3bkbu}} with default parameters and only change the number of topics.

For \textbf{RAG-topic}, which is modified from RAG\footnote{\url{https://tinyurl.com/mstamtct}} while keeping the default parameters, we have a shared query encoder $BERT_q(H)$, BART-generator and separate document encoders for each cluster \textit{i}, $BERT^i_d(p)$. 

We use a pre-trained bi-encoder from DPR\footnote{\url{https://tinyurl.com/mtdeta2w}} \cite{dpr} to initialize our encoders and create the index for each cluster in the knowledge base. Then, using the retrieval objective from DPR, we fine-tune the $BERT^i_d(p)$ and $BERT_q(H)$ using the training examples related to the $i^{th}$ cluster. Specifically, in the training data (of either MultiDoc2Dial or KILT-dialogue), if the gold knowledge of a training instance is in cluster \textit{i}, we put it into the training set for $BERT^i_d(p)$. In other words, in the fine-tuning process, each $BERT^i_d(p)$ will have a separate training set that includes only “questions” that require knowledge in cluster \textit{i}. 
After this step, we use the trained document encoders to create the fixed document index, the trained document encoders are also fixed now (non-parametric).

We continue finetuning the query encoder $BERT_q(H)$ and BART generator using all training data (either KILT-dialogue or MultiDoc2Dial) with the new retrieval results from the non-parametric retrievers.  We modify the retriever of RAG  to get the top-K passages as described in Section \ref{sec:method}. 

For \textbf{RAG-context-topic}, we modify the code provided by \citet{our-sigdial} in the same way we modify the RAG model to create \textbf{RAG-topic}, with the new dialogue history as the query during training and inference.

For \textbf{ChatGPT}, we use GPT-4 (8k context) with max\_tokens = 100, temperature = 0.5 and other parameters set as default. The following prompt is used, where \{Provided Knowledge\} is the web page containing the top-1 passage from the retriever.

\fbox{
\begin{minipage}{6.5cm}
Using the given knowledge    

\{Provided Knowledge\}

    Complete the dialogue with <system> as your role:
    
    <user>: ...
    
    <system>: ...
    
    [...]

    <user>: ...
\end{minipage}
}
We choose K=10 in all of our experiments as the baseline RAG model uses the top-10 retrieved passages for generation.

All models were trained on one RTX 3090 card.

\section{Experiment Setup}
\label{sec:experiment}
We use two \textbf{datasets} of knowledge-grounded dialogues for this study. 
\textbf{MultiDoc2Dial} \citep{multidoc2dial} consists of around 4800 
information-seeking conversations, grounded in 488 documents from 4 domains.
\textbf{KILT} \citep{kilt} is a 
dataset designed for knowledge-intensive tasks, grounded in Wikipedia. 
We only use its dialogue subtask (\textbf{KILT-dialogue}), in which one speaker must ground their utterances in a specific knowledge sentence, chosen from a Wikipedia page.
For consistency, we use  \textit{passage} to refer to the knowledge text spans we want to retrieve for response generation. 
Data examples and statistics are in Appendix \ref{app:example}, while an example comparing RAG-topic with RAG given a history from KILT-dialogue is in Appendix \ref{app:example_2}.

Because the number of topics T is a vital hyperparameter, having a way to pick T is crucial. To test the performance sensitivity of T, we experiment with different values of T and report the topic coherence scores and passage retrieval performance. To evaluate the quality of topics from topic model (topic coherence), we follow the authors of our CTM model \citep{ctm} and use external word embeddings topic coherence \citep{topic-coherence}. The evaluation metric for passage retrieval is Recall at 5 (R@5), which answers the question: out of all relevant passages, how many of them are included in the top-5 retrieved passages.

For \textbf{downstream evaluation} to compare to other baselines,
following KILT \cite{kilt}, we use page-level Precision at 1 (P@1) to report the final retrieval performances, which is the percentage of correct pages among the top-1 retrieved pages. 
For generation results, we use unigram-$\mathrm{F_1}$ score between the generated and gold responses. We also use KILT-$\mathrm{F_1}$, which only awards points when the gold-knowledge page is retrieved.

For both datasets, we compare our proposed {\bf RAG-topic} model to a baseline {\bf RAG} model. 
For MultiDoc2Dial, we also develop a  \textbf{RAG-context-topic} model to evaluate whether RAG-topic  can  add value to a prior model developed for this corpus (which we call \textbf{RAG-context}), which focused  on 
the dialogue history rather than the knowledge base \cite{our-sigdial}. 
 The base RAG-context approach has an algorithm and predictive modules to form the dialogue history (input to RAG), based on an assumption that including only turns grounded in the same document as the current turn provides a better input query.
Finally, for KILT-dialogue, we use two baselines from KILT \cite{kilt}, RAG and \textbf{BART+DPR}, which simply concatenates the retrieved passage from DPR to the dialogue history as input to BART.
\begin{table*}[t!]
\centering
\begin{tabular}{l|c|cccccccccc}
\multirow{2}{*}{} & \multicolumn{10}{c}{Number of Topics (T)} \\
&1&2&3&4&5&6&7&8&9&10\\
\hline
Topic coherence & 0.31 & 0.25 & 0.29& \underline{\textbf{0.38}} & \textbf{0.35} &0.33 & \textbf{0.35} & 0.29 &0.27& 0.22  \\
\hline 
RAG-topic / Validation  & 71.7 & 72.0 & \textbf{72.1} & \textbf{72.5} & \underline{\textbf{72.9}} & 71.1 & 71.3 & 71.9 & 68.0 &67.5\\
RAG-context-topic / Validation & 72.0 & 72.1 & \textbf{72.2} & \textbf{72.6} & \underline{\textbf{72.7}} & 71.1& 70.1& 71.8 & 71.3& 69.8\\
\hline 
RAG-topic / Test & 72.5 & 72.2 & 72.5 & \textbf{73.3} & \underline{\textbf{73.7}} & 71.5 & 70.9 & 72.3 & 68.3 &68.4\\
RAG-context-topic / Test & 72.8 & \textbf{72.9} & \textbf{72.9} & \textbf{73.2} & \underline{\textbf{74.4}} & 71.5& 71.7& 72.8 & 70.5& 70.1\\
\end{tabular}
\caption{Retrieval Results (R@5) on Test and Validation data of \textbf{MultiDoc2Dial} (average of 3 runs). \textbf{Bolded} results are significantly better than those in the same row with T=1 (no topic modeling) in a pairwise t-test (p < 0.05). The best result of each row is \underline{underlined}. }
\label{tab:ret_multi_t}
\end{table*}

\begin{table*}[t!]
\centering
\begin{tabular}{l|c|cccccccccc}
\multirow{2}{*}{} & \multicolumn{10}{c}{Number of Topics (T)} \\
&1&2&3&4&5&6&7&8&9&10\\
\hline
Topic coherence & 0.12 & \textbf{0.16} & \textbf{0.22}& \textbf{0.34} & \textbf{0.35} & \textbf{0.37} & \textbf{0.27} & \textbf{0.33} &\underline{\textbf{0.38}}& \textbf{0.36}  \\
\hline
RAG-topic / Validation & 36.3 & 36.2 & \textbf{38.5} & \underline{\textbf{40.1}}  & \textbf{38.0} & \textbf{38.7} & 30.6 & 30.3 & 25.3 & 23.5\\
RAG-topic / Test & 37.5 & 34.8 & 35.3 & \underline{\textbf{39.9}} & \textbf{39.4} & 39.7 & 31.6 & 30.9 & 26.3&24.7\\
\end{tabular}
\caption{Retrieval Results (R@5) of RAG-topic on Validation and Test data of \textbf{KILT-dialogue} (average of 3 runs) with the same annotation as Table \ref{tab:ret_multi_t}.}
\label{tab:ret_kilt_t}
\end{table*}
We also use \textbf{ChatGPT}  
as 
a retrieval-free baseline for both datasets, 
as well as use it as a response generator given the required knowledge. The knowledge source can be knowledge pages retrieved from a model (\textbf{+DPR\footnote{In KILT-dialogue, the non-ChatGPT counterpart is BART+DPR, but we only need DPR for retrieval.}}, \textbf{+RAG}, \textbf{+RAG-topic}, \textbf{+RAG-context}, \textbf{+RAG-context-topic}) or the \textbf{golden knowledge} provided in the datasets.
For example, \textbf{ChatGPT+RAG} means we feed the retrieved knowledge from the RAG model to ChatGPT's prompt to get the response.

Due to the randomness of the models (e.g. dropout from CTM training), we run each experiment 3 times and report the average results. 


\section{Results}
Tables \ref{tab:ret_multi_t} and \ref{tab:ret_kilt_t} show the \textbf{passage retrieval results} with various numbers of topics (T) for the 2 tested datasets with topic coherence scores reported in the first row of each table. 
Although our models can outperform the baseline counterparts with no topic modeling (T = 1) with the right choices of T (e.g., T = 4 or 5), certain Ts can yield lower results compared to the baselines (e.g., T = 10).
The results also show that the best T is consistent among the same dataset but different across 
 datasets (i.e., T = 5 for MultiDoc2Dial and T = 4 for KILT-dialogue). 
In contrast, for both datasets,  higher scores in topic coherence do not necessarily lead to higher retrieval results. 
Therefore, we suggest using the validation set and choosing T that achieves the highest R@5 to find the optimal T for each dataset. We then use the optimal T of each dataset to perform the downstream evaluation (i.e., T = 5 for MultiDoc2Dial and T = 4 for KILT-dialogue).
The top keywords of the clusters are in Appendix \ref{app:keywords}.

Tables \ref{tab:ret_multi} and \ref{tab:ret_kilt} show the \textbf{
page retrieval results} for MultiDoc2Dial and KILT-dialogue,
respectively. Our models significantly outperform the baseline counterparts with no topic modeling.
For MultiDoc2Dial, we witnessed an increase of 2.71 points from RAG to RAG-topic, and 4.76 points from RAG-context to RAG-context-topic. 
For KILT-dialogue, performance significantly increases by 5.46 points from RAG to RAG-topic.
These results indicate an improvement in retrieval performances when topic modeling is used in our proposed way. 

\begin{table}[t!]
\centering
\begin{tabular}{l|l}
Model & P@1\\
\hline
RAG & 64.61 \\
RAG-topic (ours) & 67.32* \\
RAG-context & 67.55 \\
RAG-context-topic (ours) & \textbf{72.31}* \\
\end{tabular}
\caption{
Retrieval results on MultiDoc2Dial (T = 5 for our models) with the best result \textbf{bolded}. Results with * are statistically significant (pairwise t-test, p < 0.05) compared to its non-topic counterpart in the prior row.}
\label{tab:ret_multi}
\end{table}

\begin{table}[t!]
\centering
\begin{tabular}{l|l}
Model & P@1\\
\hline
BART+DPR & 25.48* \\
RAG & 57.75 \\
RAG-topic (ours) & \textbf{63.21}* \\
\end{tabular}
\caption{
Retrieval results on KILT-dialogue (T = 4 for our models) with the best result \textbf{bolded}. Results with * are statistically significantly different (p < 0.05) compared to RAG.}
\label{tab:ret_kilt}
\end{table}

\begin{table}[t!]
\centering
\begin{tabular}{l|l|l}
Model & $\mathrm{F_1}$ & KILT-$\mathrm{F_1}$\\
\hline
RAG & 41.1 & 30.71\\
RAG-topic (ours) & 41.3* & 34.46*\\
RAG-context & 41.2 & 32.93\\
RAG-context-topic (ours) & \underline{42.1*} & \underline{36.21*}\\
\hline
ChatGPT & \textit{35.8} & - \\
\hspace{2mm} + RAG & \textit{44.5*} & 36.50*\\
\hspace{2mm} + RAG-topic & \textit{47.6*} & \textit{38.12}*\\
\hspace{2mm} + RAG-context & {46.9*} & 38.03*\\
\hspace{2mm} + RAG-context-topic & \textit{\textbf{49.3}*} & \textit{\textbf{39.81}*}\\
\hspace{2mm} + golden knowledge & \textit{55.2} & \textit{42.13}\\
\end{tabular}
\caption{Generation results on MultiDoc2Dial (T = 5 for our models). For ChatGPT, the  `+' part is only the \textit{knowledge retrieval result} from the mentioned model. Best non-ChatGPT results are \underline{underlined} and  best overall results (not using golden knowledge) are \textbf{bolded}. For RAG 
models (first 4 rows),  * indicates statistical significance (p < 0.05) compared to  equivalent non-topic model (one row above). 
For ChatGPT-based models,  * indicates significance (p < 0.05) compared to the non-ChatGPT version (first four rows),
\textit{italic} indicates significance compared to one row above.} 
\label{tab:gen_multi}
\end{table}

\begin{table}[t!]
\centering
\begin{tabular}{l|l|l}
Model & $\mathrm{F_1}$ & KILT-$\mathrm{F_1}$\\
\hline
BART+DPR & 15.19* & 4.37*\\
RAG & 13.19 & 9.05\\
RAG-topic (ours) & \underline{15.25}* & \underline{11.46}*\\
\hline
ChatGPT & \textit{16.12} & - \\
\hspace{2mm} + DPR & \textit{17.63}* & 11.97*\\
\hspace{2mm} + RAG & \textit{18.21}* & \textit{12.07}*\\
\hspace{2mm} + RAG-topic & \textit{\textbf{19.46}}* & \textit{\textbf{15.41}}*\\
\hspace{2mm} + golden knowledge & \textit{22.39} & \textit{18.72}\\
\end{tabular}
\caption{Generation results on KILT-dialogue (T = 4).
For the first 3 rows, * indicates statistical significance  (p < 0.05) compared to RAG. Annotation for ChatGPT-based models are the same as Table \ref{tab:gen_multi}.}
\label{tab:gen_kilt}
\end{table}

Table \ref{tab:gen_multi} shows the \textbf{response generation results} on MultiDoc2Dial. 
The first 4 rows show that our topic-based RAG 
models have 
significantly higher scores compared to their related RAG baselines (row above). 
Although the $\mathrm{F_1}$ increases are small 
the increases in KILT-$\mathrm{F_1}$ are larger 
. The bottom of the table shows that without any knowledge, ChatGPT performs very poorly 
(35.8). However, when 
knowledge is provided, ChatGPT generates responses with significantly higher  $\mathrm{F_1}$ and KILT-$\mathrm{F_1}$ 
compared to the original RAG-based versions (same retriever, different generator) in the first four rows.
We hypothesize that this dataset focuses on information-seeking conversations, so it is hard to provide the response without relevant information.

In Table \ref{tab:gen_kilt}, we report the generation results on the KILT-dialogue dataset. The first 3 rows show that RAG-topic achieves the highest scores for both metrics. Although the gain in $\mathrm{F_1}$ is marginal compared to BART+DPR, the KILT-$\mathrm{F_1}$ score is more than double (11.46 versus 4.37). Even without external knowledge, ChatGPT outperforms the three retrieval-based models. There are two reasons we can think of for this behavior. First, Wikipedia's knowledge is already built in ChatGPT internally during training. Second, this dataset is more chitchat-oriented so the response only needs to relate to the latest topic and is less strictly restricted to a specific piece of knowledge. When external knowledge is given to ChatGPT, we observe the same behavior as in MultiDoc2Dial. Specifically, given the same knowledge retrieved from a model (+DPR, +RAG or +RAG-topic), ChatGPT generates responses with higher $\mathrm{F_1}$ and KILT-$\mathrm{F_1}$ scores than its original versions. A brief analysis of the relation between the length of the dialogue history and the generation performance is in Appendix \ref{app:length}.

In general, models with higher P@1 have higher $\mathrm{F_1}$ and KILT-$\mathrm{F_1}$
\footnote{Exceptions are BART+DPR vs. RAG (KILT-dialogue) and RAG-topic vs. RAG-context (MultiDoc2Dial).}. 
Models using golden knowledge
achieve the highest results. When only retrieved knowledge is used, ChatGPT with the best retriever always wins (+RAG-context-topic for MultiDoc2Dial and +RAG-topic for KILT-dialogue).
This suggests that better retrieval leads to better generation. 

\section{Conclusion}

In this work, we proposed a method that utilizes topic modeling on the knowledge base to improve the performance of RAG-based models. 
Our approach uses topic modeling to cluster the knowledge base, build a separate document encoder for each cluster, and uses the topic distribution weights to calculate similarity scores.
Additionally, we experiment with ChatGPT to see its performance with and without external knowledge.
We observe that using the validation set to find the optimal number of topics is a reliable approach.
Overall, our RAG-based models achieve improvement in both retrieval and generation, and compliment with models focusing on building a better  dialogue history representation.
We also find that ChatGPT can take advantage of the improved retrieval performance to yield even higher generation results. 
ChatGPT does not perform very well without external knowledge, but it is superior when knowledge is provided, obtaining higher results given the same knowledge compared to RAG-based models.
Future plans include utilizing multi-task training with similar knowledge-intensive tasks and integrating the knowledge-retrieving process into a pipeline of large language models such as ChatGPT.

\section{Limitations}
One major limitation of our approach is that the computational requirement is proportional to the number of topics T as we need to retrieve from each knowledge base cluster to get the final top-K. 
For each new topic, we need an additional document encoder (BERT - 110M parameters), which results in 110M x (T-1) parameters more than the original RAG model (where T is the number of topics ).
Therefore, this method does not scale well if the optimal T is large. Additionally, for generation results, this work relies solely on automatic metrics and lacks human evaluation. The lack of diversity of open-source LLMs such as Llama2 or Vicuna makes the findings less generalizable.

\section*{Ethical Considerations}
Although this work focuses on knowledge retrieval performance (e.g. finding the correct knowledge passages as frequently as possible), other aspects of accuracy should be considered, especially in systems that provide information to the user. For example, for a healthcare application, giving the user wrong information is more dangerous than generating an irrelevant response, but both cases are considered equally failed instances when training/testing for most models. Since no NLP/AI model is perfect, depending on the application, further regulation is needed to prevent misinformation.

\section*{References}
\bibliographystyle{lrec-coling2024-natbib}
\bibliography{lrec-coling2024-example, custom}

\appendix

\section{Datasets}
\label{app:example}
\begin{figure*}[t!]
\includegraphics[width=\textwidth]{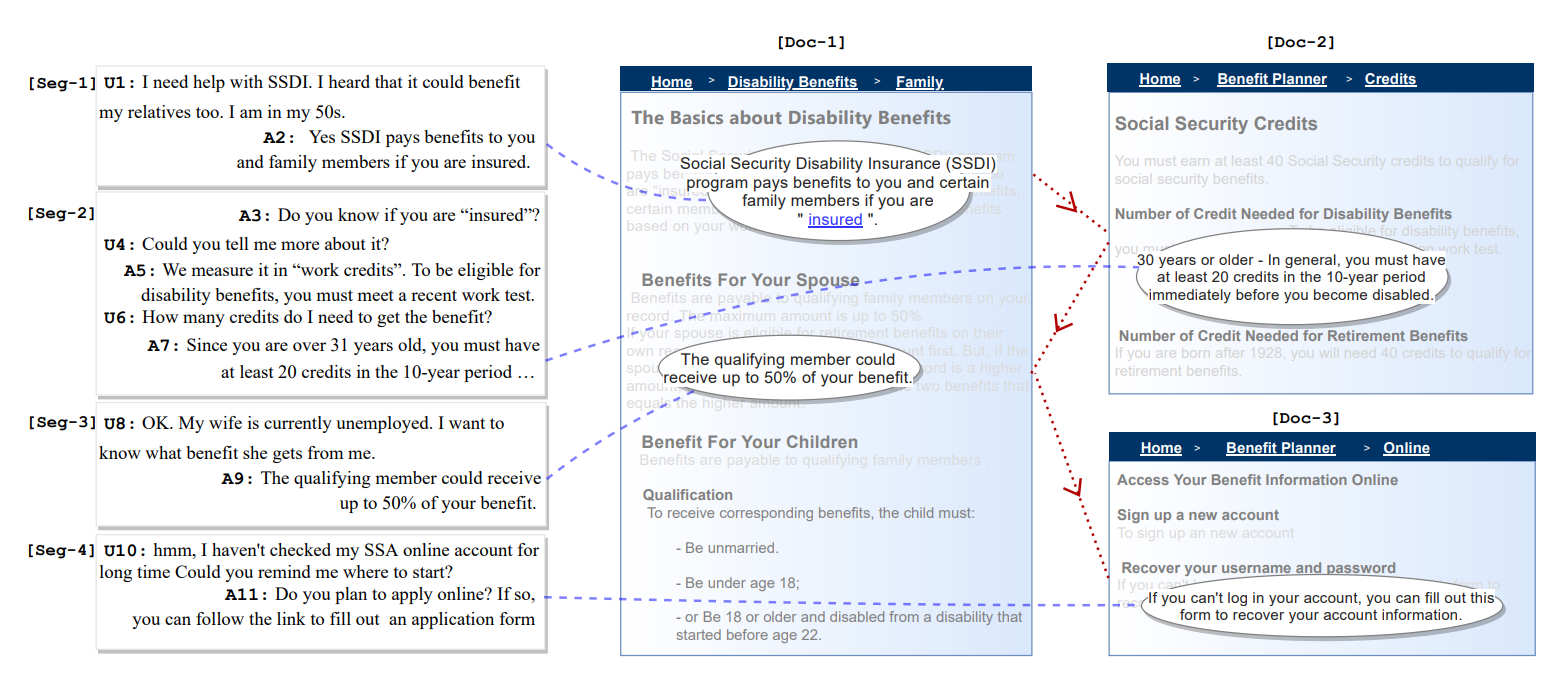}
\caption{An example dialogue from MultiDoc2Dial borrowed from \citet{multidoc2dial}. The conversation (on the left) is grounded in 3 documents Doc-1, Doc-2, and Doc-3. Each dialogue segment indicates that all turns within it are grounded in the same document (e.g., A3 to A7 in Seg-2 are all grounded in Doc-2). A dialogue
turn and its corresponding relevant span in a document are connected by a blue dashed line. The red dotted lines with arrows show the dialogue flow shifts among the grounding
documents through the conversation (e.g., Doc-1 $\rightarrow$ Doc-2 $\rightarrow$ Doc-1 $\rightarrow$ Doc-3).}
\label{fig:multidoc2dial}
\end{figure*}

\begin{figure*}[t!]
\includegraphics[width=\textwidth]{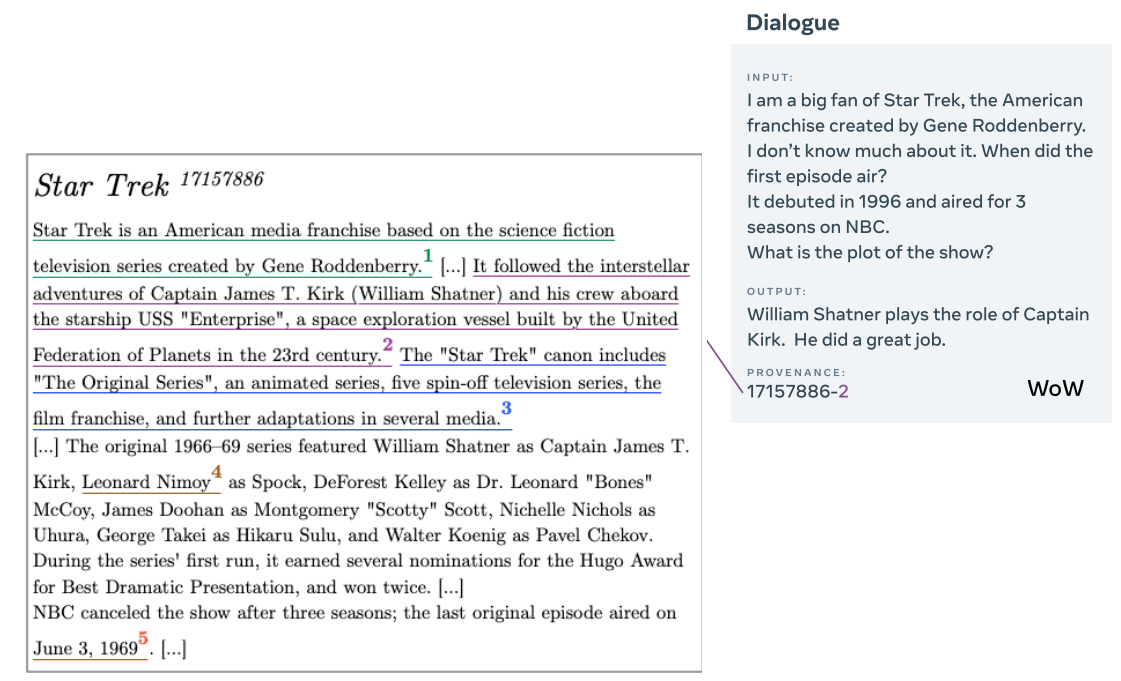}
\caption{An example dialogue from KILT-dialogue borrowed from \citet{kilt}. Two speakers talk about a given topic (e.g., Star Trek) grounded in a Wikipedia page.  }
\label{fig:wow}
\end{figure*}

Figures \ref{fig:multidoc2dial} and \ref{fig:wow} show one example each from our two datasets, MultiDoc2Dial and KILT-dialogue, respectively. 

The sizes of training, validation and test set of the two datasets we used can be seen in Table \ref{tab:data_size}. For KILT-dialogue, since the gold answer of the original test set is not released, we use the original validation set as our test set (3054 items). We then use 3054 out of the 63734 instances in the original training set as our validation set to find the optimal T. As a result, our training set consists of 60680 instances.

\begin{table}[ht!]
    \centering
    
    \begin{tabular}{lrrr}
    \hline
        Dataset & Train & Validation  & Test \\
        \hline
        MultiDoc2Dial & 3,474 & 661 & 661 \\
        KILT-dialogue & 60680 &  3,054 & 3054\\
    \end{tabular}
    \caption{Dataset Statistics}
    \label{tab:data_size}
\end{table}

\begin{table*}[t!]
    \centering
    
    \begin{tabular}{p{0.13\linewidth} | p{0.09\linewidth}  p{0.68\linewidth}}
        \multirow{5}{*}{KILT-Dialogue} & \multicolumn{2}{c}{Number of Topics (T) = 4} \\
        \cline{2-3}
         &Cluster 1 & east, west, south, river, north, state, area, city, district, center\\
        &Cluster 2 & rock, band, records, music, song, album, team, record, club, studio\\
        &Cluster 3 & story, fiction, characters, book, disney, novel, film, episode, films, comic \\
        &Cluster 4 & pain, bon, Canberra, rutgers, blocked, khalil, edmonton, capitals, auckland, auburn \\
        \hline
        \multirow{6}{*}{MultiDoc2Dial} & \multicolumn{2}{c}{Number of Topics (T) = 5} \\
        \cline{2-3}
         &Cluster 1 & car, dmv, vehicle, plate, license, driver, toll, registration, insurance, hearing\\
        &Cluster 2 & benefit, social, disabled, number, income, retirement, document, children, child, security\\
        &Cluster 3 & student, aid, school, apply, scholarship, college, aids, program, grant, loans\\
        &Cluster 4 &  va, status, appeal, account, claim, evidence, review, compensation, deposit, allowance\\
        &Cluster 5 &  test, benefits, address, registrations, information, website, programs, accounts, online, office\\
    \end{tabular}
    \caption{Top 10 words for each cluster of the knowledge base of KILT-dialogue and MultiDoc2Dial}
    \label{tab:cluster_keywords}
\end{table*}

\begin{table*}[t!]
    \centering
    
    \begin{tabular}{p{0.22\linewidth} | p{0.33\linewidth} | p{0.33\linewidth}}
    \hline
        \multicolumn{3}{c}{Dialogue history}    \\ \hline
        \multicolumn{3}{l}{Speaker 1: the Draco lizard is so cool they can glide from trees }  \\
        \multicolumn{3}{l}{Speaker 2: Lizards are just cool in general but i havent heard of that one before } \\
        \multicolumn{3}{l}{Speaker 1: have you heard of Draco Malfoy?} \\
        \hline \hline
        Model & \textbf{RAG} & \textbf{RAG-topic} (T = 4) \\ \hline
        Topic distribution & w = (1.00) &  w = (0.21, 0.09, 0.55, 0.15) \\
        \hline
        Retrieved passage (Top-1) & Members of Draco are primarily arboreal, inhabiting tropical rainforests, and are almost never found on the forest floor & Draco Lucius Malfoy is a character in J. K. Rowling's "Harry Potter" series. \\
        \hline
        Generated response & Yes, you can find them in tropical rainforests. & Yes, he is a character in harry potter series.\\
    \end{tabular}
    \caption{An example from KILT-dialogue in which our proposed RAG-topic successfully retrieved a relevant knowledge passage while the original RAG failed to do so for the same given dialogue history. For RAG-topic,  vector w represents the topic distribution of the four clusters in Table \ref{tab:cluster_keywords} from the dialogue history. }
    \label{tab:our_example}
\end{table*}

\section{Examples of Retrieved Passages and Response Generation}
\label{app:example_2}
In Table \ref{tab:our_example}, we show the top-1 retrieved passage and generated response from RAG and RAG-topic for a given dialogue history in KILT-dialogue. The topic distribution weights from CTM helped guide the search to Cluster 3, which contains knowledge about novels and films, to find a relevant knowledge passage. On the other hand, the original RAG model found an irrelevant knowledge passage and generated an inappropriate response.

\section{Themes Among the Clusters}
\label{app:keywords}
Table \ref{tab:cluster_keywords} shows the list of keywords of each cluster from the knowledge base of KILT-dialogue when the number of topic T for CTM is set as 4 and MultiDoc2Dial when T is set as 5. 
Generally, there are “themes” among these clusters. For example, in KILT-Dialogue, cluster 1 is related to geography, cluster 2 is about music, cluster 3 is about novels and film. For MultiDoc2Dial, the first 4 clusters are quite representative of the 4 domains in the dataset: Department of Motor Vehicles (dmv), Social Security Affair (ssa), Student Aid (sa) and Veteran Affair (va). The last cluster is a mixture of the information across 4 domains.

\section{Relations between Dialogue History Length and Performance}
\label{app:length}
Table \ref{tab:length} shows the Pearson correlation values between the length of the dialogue history (number of tokens) and generation performance (F1) on two datasets. Generally, there are negative correlations between the two variables, indicating that the performance decreases when the dialogue history is longer. This is reasonable as not all information in the dialogue history is relevant to the current turn and the redundancy can create noise for the retrieval process. We also observe that our proposed approaches (RAG-topic and RAG-context-topic) help mitigate this negative relation because their absolute Pearson correlation values are smaller compared to RAG. Especially in MultiDoc2Dial, RAG-context-topic can filter out irrelevant turns in the history, and thus there might be less noise in the selected dialogue history used as a query for the retrieval process.

\begin{table*}[ht!]
    \centering
    
    \begin{tabular}{lcc}
    \hline
         & MultiDoc2Dial & KILT-dialogue\\
        \hline
        RAG & -0.35* & -0.24* \\
        RAG-topic & -0.28* & -0.20*\\
        RAG-context-topic & -0.19* & n/a\\
    \end{tabular}
    \caption{The Pearson correlation values between the length of the dialogue history (number of tokens) and generation performance (F1). The columns represent the different datasets
and the rows represent the different models. * indicates p < 0.05 in a two-tailed t-test.}
    \label{tab:length}
\end{table*}
\end{document}